\documentclass[aps,11pt,showpacs,showkeys]{revtex4}
\begin{document}
\title{Criticality in self-dual sine-Gordon models}
\author{P. Lecheminant}
\email{Philippe.Lecheminant@ptm.u-cergy.fr}
\affiliation{Laboratoire de Physique Th\'eorique et
Mod\'elisation, CNRS ESA 8089,
Universit\'e de Cergy-Pontoise, 5 Mail Gay-Lussac, Neuville sur Oise,
95301 Cergy-Pontoise Cedex, France}
\author{Alexander O. Gogolin}
\affiliation{
Department of Mathematics, Imperial College,
180 Queen's Gate, London SW7 2BZ, United Kingdom}
\author{Alexander A. Nersesyan}
\affiliation{The Abdus Salam International Centre for Theoretical Physics,
P.O.Box 586, 34100, Trieste, Italy}
\begin{abstract}

We discuss the nature of criticality in
the $\beta^2 = 2 \pi N$ self-dual extention of the sine-Gordon
model. This field theory is related
to the two-dimensional classical XY model
with a N-fold degenerate symmetry-breaking field.
We briefly overview the already studied cases $N=2,4$
and analyze in detail the case $N=3$ where
a single phase transition in the three-state
Potts universality class is expected to occur.
The Z$_3$ infrared critical properties
of the $\beta^2 = 6 \pi$ self-dual sine-Gordon
model are derived using two non-perturbative approaches.
On one hand, we map the model onto  an integrable deformation of
the Z$_4$ parafermion theory.
The latter is known to flow to a massless Z$_3$ infrared fixed point.
Another route is based on the connection with a
chirally asymmetric, su(2)$_4$ $\otimes$ su(2)$_1$
Wess-Zumino-Novikov-Witten model with
anisotropic current-current interaction, where we explore
the existence of a decoupling (Toulouse) point.

\end{abstract}
\maketitle

%%%%%%%%%%%%%%%%%%%%%%%%%%%%%%%%%%%%%%%%%%%%%%%%%%%%%%%%%%%%%%%%%%%%%%%%%%%%%
\section{Introduction}

The emergence of
a non-trivial criticality in
a conformal field theory (CFT)
perturbed by several competing relevant operators 
has attracted much interest in recent years 
in the context of two-dimensional statistical mechanics 
or one-dimensional quantum systems \cite{mussardo,fgn,wang,wangbis}.
When acting separately, each perturbation yields a massive field theory,
but the interplay between them may give rise to a second-order phase
transition at intermediate coupling.
The lack of integrability in such models
and the inapplicability of perturbation theory in this situation
makes it difficult to analyse the vicinity of the
intermediate fixed point that separates physically different,
strong-coupling massive phases. However, a good understanding of such
criticality can still be reached in some special cases.
A concrete example is the double-frequency sine-Gordon model (DSG)\cite{mussardo},
which is the Gaussian model of a bosonic field $\Phi$
perturbed by two scalar vertex operators with the ratio of their scaling
dimensions equal to 4. An Ising critical point occurs when
the two perturbations cannot be minimized simultaneously,
i.e. when the two massive phases generated by each perturbation cannot
be connected by a continuous path in the parameter space of the model.
The nature of the resulting phase transition has been clarified
by general arguments concerning the excitation spectrum of
the DSG model \cite{mussardo}. A non-perturbative treatment of
the Ising critical point was proposed in Ref. \cite{fgn}. This approach
is based on a quantum lattice version of the model that enables one to clearly
identify the fast and slow degrees of freedom of the problem.
The existence of the Ising criticality has also been confirmed using
the truncated conformal space approach \cite{bajnok}.
The DSG model has interesting applications in one-dimensional
quantum magnetism (a spontaneously dimerized spin-1/2 Heisenberg chain
in a staggered magnetic field \cite{fgn}, a two-leg spin-1
ladder \cite{allenspin1}) and one-dimensional models of interacting electrons
(the half-filled Hubbard model with alternating chemical potential \cite{fgn,fgnlett},
the quarter-filled electron system with dimerization \cite{tsuorignac}).
\medskip

In this paper, we investigate the critical properties of another deformation of the
sine-Gordon model obtained by adding a second relevant vertex operator
which depends on the field $\Theta$ dual to the field $\Phi$.
The action of this model is given by
\begin{equation}
{\cal S} = \frac{1}{2} \int d^2 {\bf r} \; \left(\partial_{\mu} \Phi\right)^2
+ g \int d^2 {\bf r} \; \cos\left(\beta \Phi\right)
+ {\tilde g} \int d^2 {\bf r} \; \cos\left({\tilde \beta} \Theta\right).
\label{actiondep}
\end{equation}
Notice that the two perturbations in Eq. (\ref{actiondep}), with scaling
dimensions $\Delta_g = \beta^2/4\pi$ and $\Delta_{\tilde g} =
{\tilde \beta}^2/4\pi$, are mutually nonlocal and, therefore, cannot be
minimized simultaneously. When $\Delta_g\ne \Delta_{\tilde g} < 2$, the
low-energy physics is governed by the most relevant operator, and the
problem effectively reduces to the standard sine-Gordon model, either for
the field $\Phi$ ($\Delta_g < \Delta_{\tilde g}$) or for the 
dual field $\Theta$
($\Delta_{\tilde g} < \Delta_g$). In this case the resulting infrared (IR) 
theory is fully massive.
However, the most interesting situation arises when
the two perturbations are both relevant and have the same scaling dimension:
$\beta = {\tilde \beta}$, $\Delta_g < 2$. Then the competition between these
two antagonistic terms can lead to a non-perturbative critical point (or line)
at a finite coupling. Note that at $\beta = {\tilde \beta}$ and $g = {\tilde g}$
the action (\ref{actiondep}) becomes invariant under the duality transformation
$\Phi \leftrightarrow \Theta$. In what follows, such model will be referred to
as the self-dual sine-Gordon (SDSG) model. Its self-duality opens a possibility
for the existence of a critical point. From the renormalization
group (RG) point of view, the SDSG
model will then be characterized by a massless flow from the 
ultraviolet (UV) Gaussian fixed point with
central charge $c_{UV}=1$ to a conformally invariant IR fixed point
with a smaller central charge $c_{IR} < 1$ according to the Zamolodchikov's
c-theorem \cite{ctheorem}. 
The underlying CFT will necessarily be a member of the minimal
model series, since these are the only unitary CFTs
with central charge $c < 1$ \cite{friedan}.
\medskip

Much insight on the critical properties of the SDSG model can be gained by
considering the related two-dimensional classical XY model with a
N-fold symmetry-breaking field. The lattice Hamiltonian of this system reads:
\begin{equation}
{\cal H}_N = - J \sum_{<{\bf r}, {\bf r}^{'}>}
\cos\left(\theta_{{\bf r}} - \theta_{{\bf r}^{'}}\right)
+ h \sum_{{\bf r}} \cos\left(N \; \theta_{{\bf r}}\right),
\label{xyanisoham}
\end{equation}
where $\theta_{{\bf r}}$ is the angle of the unit-length
rotor at site ${\bf r}$ of a square lattice, and the symbol
$<{\bf r}, {\bf r}^{'}>$ indicates summation over the nearest-neighbor sites.
The last term in Eq. (\ref{xyanisoham}) breaks the continuous O(2) symmetry
of the XY model down to a discrete one, Z$_N$ ($N=2,3,...$).
\medskip

The model (\ref{xyanisoham}) has a long history and has been extensively
studied. At $N \ge 5$ there are two phase transitions as a
function of the temperature \cite{jose}.
For $T < T_{c1}$ the symmetry breaking field is dominant, and the system
occurs in the broken symmetry phase in which one of the global directions
$\theta = 2\pi n/N, n=0,1,..,N-1$ is preferred. At $T > T_{c2}$ the system
is in a paramagnetic phase with unbroken Z$_N$ symmetry and short-range order.
In the temperature window $T_{c1} < T < T_{c2}$ the system is in
a Gaussian, XY-like phase with power-law correlations.
However, for $N \le 4$ such intermediate massless phase
is absent, and  a direct transition from
the ``ferromagnetic'' phase to the paramagnetic one takes place.
The type of the emerging criticality is determined by the symmetry of the model.
Thus the criticality should belong to the Ising and three-state Potts universality
classes at $N=2$ and $N=3$, respectively. A perturbative RG analysis predicts
that the $N=4$ criticality is characterized by nonuniversal, continuously varying
exponents. Finally, in the $N=1$ case, the O(2) symmetry is broken at all
temperatures. Similar conclusions have been reached for the Z$_N$ clock model
\cite{elitzur} which is equivalent to the Hamiltonian (\ref{xyanisoham})
in the large-$h$ limit. The critical properties of the clock models
have been further investigated by means of series expansions \cite{elitzur,hamer}
and exact diagonalization calculations on finite samples \cite{roomany,bonnier}.
\medskip

Close to the transition point,
all universal (long-distance) properties of the classical lattice model
(\ref{xyanisoham}), or the related Z$_N$ clock model with
$2 \le N \le 4$, are adequately described
within a continuum description based on the effective action \cite{wiegman}:
\begin{equation}
S = \frac{K}{2} \int d^2 {\bf r} \; \left(\partial_{\mu} \Phi\right)^2
+ g \int d^2 {\bf r} \; \cos\left(2 \pi \Theta\right)
+ h \int d^2 {\bf r} \; \cos\left(N \Phi\right),
\label{xyancont}
\end{equation}
where $K$ is the stiffness of the Bose field $\Phi$.
The first two terms in Eq. (\ref{xyancont}) constitute
the effective action of the O(2)-symmetric XY model, with the cosine of the dual
field accounting for topological vortices (for a review see e.g.
Refs. \cite{kogut,wiegmanbis,tsvelikbook}),
whereas the last term represents the symmetry breaking perturbation.
By a simple rescaling of the Bose field,
actions (\ref{xyancont}) and (\ref{actiondep})
become identical, with $\beta = N/\sqrt{K}$
and ${\tilde \beta} = 2\pi \sqrt{K}$, and
the self-duality condition becomes
$\beta = {\tilde \beta} = \sqrt{2\pi N}$.
\medskip

In this paper, we discuss the critical properties of 
the $\beta^2 = 2\pi N$ SDSG model
\begin{equation}
{\cal S}_{\rm SDSG} = \frac{1}{2}
\int d^2 {\bf r} \; \left(\partial_{\mu} \Phi\right)^2
+ g \int d^2 {\bf r} \; \left(\cos\left(\sqrt{2\pi N} \; \Phi\right)
+ \cos\left(\sqrt{2\pi N} \; \Theta\right)\right),
\label{actionsdsg}
\end{equation}
at $2 \le N \le 4$. The fact that at $N=2$ and 4 the criticality
belongs to the Z$_N$ universality class is already well known
and can be reproduced using the standard bosonization/refermionization
techniques. For completeness, we review these cases in Section II.
To the best of our knowledge, the case $N=3$ is still lacking a consistent
non-perturbative analytical description. The main difficulty stems from the fact
that the underlying field theory, Eq. (\ref{actionsdsg}), does not
admit a simple free field representation
and, as opposed to the DSG model \cite{fgn}, it
does not suggest any clear decomposition between the
fast and slow degrees of freedom.  The resulting IR fixed point is
strongly non-perturbative in this respect. In what follows,
the three-state Potts universality class
of the criticality in the $\beta^2 = 6\pi$ SDSG model will be {\sl derived}
using two different routes.
First, we shall map the N=3 action (\ref{actionsdsg}) onto
an integrable deformation of the Z$_4$ parafermion theory \cite{fateevzam}
and exploit the existence of a massless flow from this model to
the three-state Potts criticality.
On the other hand, we shall also relate the $\beta^2 = 6\pi$ SDSG model
to a chirally asymmetric version of a Wess-Zumino-Novikov-Witten (WZNW)
model with su(2)$_4$--su(2)$_1$ current-current interaction.
That theory is integrable and, in the IR limit,
displays the properties of the so-called
chirally stabilized liquids \cite{andrei}.
From the symmetry of this IR fixed point,
the three-state Potts criticality of the $\beta^2 = 6\pi$ SDSG model
will be deduced by considering a special anisotropic version of
the above WZNW model. Finally, this approach will enable us 
to address the UV-IR transmutation 
of some fields of the $\beta^2 = 6\pi$ SDSG model. 
\medskip

The rest of the paper is organized as follows.
In section II we review the properties of the $\beta^2 = 2\pi N$
SDSG model (\ref{actionsdsg}) in the simplest cases
$N=1,2,4$. The mapping of the $\beta^2 = 6\pi$ SDSG model
onto an integrable deformation of the Z$_4$ parafermion theory
is presented in Section III. In Section IV
we study an anisotropic version of the
WZNW model with su(2)$_4$ and su(2)$_1$ current-current interaction
and address its relationship to the $\beta^2 = 6\pi$ SDSG model.
Our concluding remarks are summarized
in Section V.
The paper is  supplied with an Appendix which
provides some details on the bosonization approach to the Z$_4$ parafermion
theory.

%%%%%%%%%%%%%%%%%%%%%%%%%%%%%%%%%%%%%%%%%%%%%%%%%%%%%%%%%%%%%%%%%%%%%%%%%%%%%
\section{Review of some simple cases}

In this section, we overview the IR properties of the
$\beta^2 = 2\pi N$ SDSG model  (\ref{actionsdsg}) in the simple cases $N=1,2,4$.
The $\beta^2 = 2\pi$ SDSG model
describes a massive field theory since the related
lattice model (\ref{xyanisoham}) for $N=1$ has no magnetic phase transition.
In contrast, the $\beta^2 = 4\pi$ SDSG model
displays critical properties in the Ising universality class,
whereas the $\beta^2 = 8\pi$ case corresponds to
a Gaussian CFT with central charge $c=1$.

%%%%%%%%%%%%%%%%%%%%%%%%%%%%%%%%%%%%%%%%%%%%%%%%%%%%%%%%%%%%%%%%%%%%%%%%%%%%%
\subsection{The $\beta^2 = 2\pi$ SDSG model}

Let us first consider the case $N=1$. The lattice Hamiltonian (\ref{xyanisoham})
describes the two-dimensional classical XY model in
a magnetic field along the x-axis. The effective field theory associated with
this system is given by the $\beta^2 = 2\pi$ SDSG model whose quantum Hamiltonian
reads:
\begin{equation}
{\cal H}_{\beta^2 = 2\pi} = \frac{1}{2} \left[
\left(\partial_x \Phi\right)^2 +
\left(\partial_x \Theta\right)^2 \right] +
g\left[\cos\left(\sqrt{2\pi} \; \Phi\right)
+ \cos\left(\sqrt{2\pi} \; \Theta\right)\right].
\label{ham2pi}
\end{equation}
The fact that no magnetic phase transition takes place in the $N=1$ lattice
model (\ref{xyanisoham}) can be inferred from the relationship between the
field theory (\ref{ham2pi}) and an explicitly dimerized spin-1/2
antiferromagnetic Heisenberg chain in a staggered magnetic field along
the x-direction. The Hamiltonian of this quantum spin chain is
\begin{equation}
{\cal H} = J\sum_i {\bf S}_{i}\cdot {\bf  S}_{i+1}
+  \Delta \sum_i \left(-1\right)^i {\bf S}_{i}\cdot {\bf  S}_{i+1}
+ h \sum_i \left(-1\right)^i S_i^x ,
\label{hamlat2pi}
\end{equation}
where ${\bf S}_{i}$ is the spin-1/2 operator at the lattice site $i$.
In the low-energy limit, the standard spin-1/2 Heisenberg chain, given by
the first term in Eq. (\ref{hamlat2pi}), is described by the critical su(2)$_1$
WZNW model \cite{affleck,affleckhaldane}(with a marginally irrelevant
current-current perturbation). This model, in turn, can be bosonized
and recast as a simple Gaussian model with compactification radius
$R = 1/\sqrt{2\pi}$ of the bosonic field $\Phi$:
\begin{equation}
{\cal H}_0 = \frac{1}{2} \left[
\left(\partial_x \Phi\right)^2 +
\left(\partial_x \Theta\right)^2 \right]
\label{freebosham}
\end{equation}
(for simplicity, we have set the velocity $v=1$). In the continuum description,
the local spin density ${\bf S}(x)$ separates into the smooth and
staggered parts:
\begin{equation}
{\bf S} \left(x\right) = {\bf J}\left(x\right) +
\left(-1\right)^{x/a} {\bf n} \left(x\right),
\label{inispindens}
\end{equation}
where $a$ is the lattice spacing. The bosonized expression for the
staggered magnetization is \cite{bookboso}
\begin{equation}
{\bf n} = \frac{\lambda}{\pi a} \left[
\cos\left(\sqrt{2\pi} \ \Theta\right),
\sin\left(\sqrt{2\pi}\ \Theta\right),
-\sin\left(\sqrt{2 \pi}\ \Phi\right) \right],
\label{inspinstagboso}
\end{equation}
$\lambda$ being a nonuniversal constant.
Similarly, the dimerization operator
transforms to
\begin{equation}
(-1)^i {\bf S}_i \cdot {\bf S}_{i+1}
\to \epsilon (x) = \frac{\lambda}{\pi a} \cos \left(\sqrt{2 \pi}\ \Phi\right).
\label{dim-op}
\end{equation}
Using the expressions (\ref{inspinstagboso}) and (\ref{dim-op})
and properly fine tuning the coupling constants, one establishes
the correspondence between the Hamiltonian (\ref{hamlat2pi})
and the $\beta^2 = 2\pi$ SDSG model, Eq. (\ref{ham2pi}).
\medskip

The next step is to perform a spin rotation along the y-axis
in the Hamiltonian (\ref{hamlat2pi}), so that the staggered
magnetic field becomes applied in the z-direction.
This transformation does not affect the dimerization term
(which is SU(2) invariant) but otherwise makes the new continuum model
dependent only on a Bose field $\Phi^{\prime}$ (no dual 
field in the interaction):
\begin{equation}
{\cal H}_{\beta^2 = 2\pi}^{'} = \frac{1}{2} \left[
\left(\partial_x \Phi^{\prime}\right)^2 +
\left(\partial_x \Theta^{\prime}\right)^2 \right] +
g\left[\cos\left(\sqrt{2\pi} \; \Phi^{\prime}\right)
- \sin\left(\sqrt{2\pi} \; \Phi^{\prime}\right)\right].
\label{ham2pibis}
\end{equation}
This Hamiltonian reduces to the standard $\beta^2 = 2\pi$ sine-Gordon model
by a shift $\Phi^{\prime} \to \Phi^{\prime} - \sqrt{\pi}/4\sqrt{2}$.
The latter is a fully massive integrable field theory. Thus, there is
no quantum critical point in the phase diagram of the 
$\beta^2 = 2\pi$ SDSG model.

%%%%%%%%%%%%%%%%%%%%%%%%%%%%%%%%%%%%%%%%%%%%%%%%%%%%%%%%%%%%%%%%%%%%%%%%%%%%%
\subsection{The $\beta^2 = 4\pi$ SDSG model}

We now turn to the analysis of the $N=2$ case when the Hamiltonian
takes the form
\begin{equation}
{\cal H}_{\beta^2 = 4\pi} = \frac{1}{2} \left[
\left(\partial_x \Phi\right)^2 +
\left(\partial_x \Theta\right)^2 \right] +
g\left[\cos\left(\sqrt{4\pi} \; \Phi\right)
+ \cos\left(\sqrt{4\pi} \; \Theta\right)\right].
\label{ham4pi}
\end{equation}
This model is well known (see e.g. Ref. \cite{ogilvie}) and has
a number of applications; for instance, it appears
in the context of weakly coupled Heisenberg spin chains \cite{shelton,schulz}.
It can be exactly diagonalized even in a more general case when the two cosine
terms in Eq. (\ref{ham4pi}) have independent amplitudes,
$g$ and $\tilde{g}$. Since
each of the two perturbation has scaling dimension 1 and, given the fact that
the Hamiltonian does not possess any continuous symmetry, the model can be
refermionized by introducing two Majorana fields, $\xi^{1,2}$.
This procedure is nothing but the standard bosonization of two Ising
models \cite{zuber,truong,boyanovsky}.
The bosonization rules are given by
\begin{eqnarray}
\xi_R^{1} + i \xi_R^{2} &=& \frac{1}{\sqrt{\pi}} :\exp\left(i\sqrt{4\pi}
\Phi_R\right):, \nonumber \\
\xi_L^{1} + i \xi_L^{2} &=& \frac{1}{\sqrt{\pi}} :\exp\left(-i\sqrt{4\pi}
\Phi_L\right):,
\label{bosoising}
\end{eqnarray}
where $\Phi_{R,L}$ are the chiral components of the Bose field:
$\Phi_L = (\Phi + \Theta)/2$ and $\Phi_R = (\Phi - \Theta)/2$.
These fields are normalized according to
$\langle \Phi_L (z) \Phi_L (w) \rangle = - \ln(z-w)/4\pi$
and $\langle \Phi_R (\bar z) \Phi_R (\bar w) \rangle
= - \ln(\bar z - \bar w)/4\pi$ with $z= \tau + ix$ and
$\bar z= \tau - ix$ ($\tau$ being the Euclidean time).
In the present work, all chiral Bose fields will be normalized
in this way. One can easily check that the bosonic representation
(\ref{bosoising}) is consistent with the standard operator
product expansion (OPE) for the Majorana fields:
\begin{eqnarray}
\xi_L^a\left(z\right) \xi_L^b\left(w\right)
&\sim& \frac{\delta^{ab}}{2\pi\left(z-w\right)},
\nonumber \\
\xi_R^a  \left(\bar z\right)
\xi_R^b\left(\bar w\right)
&\sim& \frac{\delta^{ab}}{2\pi\left(\bar z- \bar w\right)}.
\label{majope}
\end{eqnarray}
Strictly speaking, the doublet of identical critical
Ising copies is not equivalent to the CFT of a free massless Dirac fermion
since the two Majorana fermions in Eq. (\ref{bosoising})
are not independent but constrained to share the same type of 
boundary conditions.
The Dirac CFT can be bosonized using a free massless scalar field
compactified on a circle with radius $R = 1/\sqrt{4\pi}$ in
our notation. In contrast, two decoupled Ising models
are described by a Bose field living on the orbifold line
with the same radius (for a review on this identification, see for instance
Refs. \cite{ginsparg,dms}). Nonetheless,
as far as the bulk properties of the
$\beta^2 = 4 \pi$ SDSG model (\ref{ham4pi}) are concerned,
this subtlety does not manifest itself and
the correspondence (\ref{bosoising}) can be safely applied.
The self-dual Hamiltonian (\ref{ham4pi}) can then be expressed in terms
of these Majorana fermions:
\begin{equation}
{\cal H}_{\beta^2 = 4\pi} = -\frac{i}{2} \sum_{a=1}^{2}\left(
\xi_R^{a} \partial_x \xi_R^{a} -
\xi_L^{a} \partial_x \xi_L^{a} \right)
+ i m \xi_R^{2} \xi_L^{2},
\label{ham4pifer}
\end{equation}
with $m = 2  \pi g$. One thus observes that the Hamiltonian
of the $\beta^2 = 4\pi$ SDSG model separates into two commuting pieces.
One of the decoupled degrees of freedom corresponds to an effective
off-critical Ising model described by the massive Majorana
fermion $\xi_{R,L}^2$, whereas
the second Majorana field $\xi_{R,L}^1$ remains massless.
(In the two-parameter model with $g \neq  \tilde{g}$ both sectors are massive:
 $m_1 = \pi (g-\tilde{g})$ and $m_2 = \pi (g+\tilde{g})$.)
The existence of a massless Majorana mode signals the Z$_2$
(Ising) criticality of the $\beta^2 = 4\pi$ SDSG model (\ref{ham4pi}).

%%%%%%%%%%%%%%%%%%%%%%%%%%%%%%%%%%%%%%%%%%%%%%%%%%%%%%%%%%%%%%%%%%%%%%%%%%%%%
\subsection{The $\beta^2 = 8\pi$ SDSG model}

Here we briefly review the remaining simple case, $\beta^2 = 8\pi$, when the
Hamiltonian has the form:
\begin{equation}
{\cal H}_{\beta^2 = 8\pi} = \frac{1}{2} \left[
\left(\partial_x \Phi\right)^2 +
\left(\partial_x \Theta\right)^2 \right] +
g\left[\cos\left(\sqrt{8\pi} \; \Phi\right)
+ \cos\left(\sqrt{8\pi} \; \Theta\right)\right].
\label{ham8pi}
\end{equation}
The self-dual interaction is now marginal, so
it is natural to expect a critical behavior. The perturbative RG approach is
applicable to this case and indicates the existence of a line of $c=1$ fixed points
\cite{jose,wiegman,wiegmanbis,giamarchi,boyanovskyrg}.
The $\beta^2 = 8\pi$ SDSG model emerges in the problem of the one-dimensional Fermi gas
with backscattering and spin-nonconserving processes \cite{truongbis}, and it also describes
critical properties of weakly coupled Luttinger chains \cite{nersluther}.
The model (\ref{ham8pi}) is also related to the quantum Ashkin-Teller model
(two identical quantum Ising chains coupled by a
self-dual interchain coupling, see e.g. Ref. \cite{fgn}).
\medskip

A simple way to clarify the nature of the criticality
of the $\beta^2 = 8\pi$ SDSGM is to
map the Hamiltonian (\ref{ham8pi}) onto
an anisotropic version of the su(2)$_1$ WZNW model with
a current-current interaction. This can be done by exploiting the fact that
the chiral currents of the su(2)$_1$ Kac-Moody (KM) algebra
have a free-field representation in terms of
a massless Bose field $\Phi$ (see Ref. \cite{ZF}):
\begin{eqnarray}
J_L^z &=& \frac{i}{\sqrt{2\pi}} \; \partial \Phi_L \nonumber \\
J_L^+ &=& \frac{1}{2\pi} \;:\exp\left( i \sqrt{8\pi}\Phi_L\right): \nonumber \\
J_R^z &=& \frac{-i}{\sqrt{2\pi}} \; {\bar \partial} \Phi_R \nonumber \\
J_R^+ &=& \frac{1}{2\pi} \;:\exp\left(-i \sqrt{8\pi}\Phi_R\right): ,
\label{su21freerep}
\end{eqnarray}
with $J_{L,R}^{\pm} = J_{L,R}^x \pm i J_{L,R}^y$.
Representation (\ref{su21freerep}) correctly reproduces the su(2)$_1$ KM algebra
($a=x,y,z$):
\begin{eqnarray}
J_{L}^{a}\left(z\right) J_{L}^{b}\left(w\right) \sim
\frac{\delta^{ab}}{8\pi^2\left(z-w\right)^2} +
\frac{i \epsilon^{a b c} J_{L}^{c}\left(w\right)}{2\pi\left(z-w\right)},
\nonumber \\
J_{R}^{a}\left(\bar z\right) J_{R}^{b}\left(\bar w\right) \sim
\frac{\delta^{ab}}{8\pi^2\left(\bar z - \bar w\right)^2} +
\frac{i \epsilon^{a b c} J_{R}^{c}
\left(\bar w\right)}{2\pi\left(\bar z - \bar w\right)},
\label{su21ope}
\end{eqnarray}
where $\epsilon^{a b c}$ is the totally antisymmetric tensor.
The Hamiltonian (\ref{ham8pi}) can then be entirely expressed in terms
of the vector currents:
\begin{equation}
{\cal H}_{\beta^2 = 8\pi} = \frac{2\pi}{3} \left( {\bf J}_L \cdot
{\bf J}_L + {\bf J}_R \cdot {\bf J}_R\right)
- 8\pi^2 g \; J_L^x  J_R^x .
\label{ham8picurr}
\end{equation}
Due to the SU(2) symmetry of the unperturbed Hamiltonian, the interaction
can alternatively be expressed in terms of the z-component of the currents.
The resulting Hamiltonian can be further bosonized using the
correspondence (\ref{su21freerep}):
\begin{eqnarray}
{\cal H}_{\beta^2 = 8\pi} &=& \frac{1}{2} \left[
\left(\partial_x \Phi\right)^2 + \left(\partial_x \Theta\right)^2 \right]
- 4\pi g \; \partial_x \Phi_L \partial_x \Phi_R \nonumber \\
&=& \frac{v}{2} \left[ K \left(\partial_x \Theta\right)^2
+ \frac{1}{K} \left(\partial_x \Phi\right)^2 \right],
\label{ham8pireboso}
\end{eqnarray}
where the role of interaction $g$ in Eq. (\ref{ham8picurr})
is exhausted by renormalization of the velocity and compactification radius:
$v^2 = 1 - 4\pi^2 g^2$ and $K^2 = (1+ 2 \pi g)/(1 - 2 \pi g)$.
We thus deduce that the $\beta^2 = 8 \pi$ SDSG model
displays Gaussian critical properties parameterized
by the exponent $K$. Correlation functions in
this model are characterized by continuously varying ($K$ dependent)
critical exponents, which is a distinctive feature
of the Luttinger-liquid universality class \cite{voit}.

%%%%%%%%%%%%%%%%%%%%%%%%%%%%%%%%%%%%%%%%%%%%%%%%%%%%%%%%%%%%%%%%%%%%%%%%%%%%%

\section{Mapping of the $\beta^2=6\pi$ SDSG model onto an
integrable deformation of the Z$_4$ parafermion theory}

Now we turn to our main problem and discuss the $\beta^2 = 6 \pi$ SDSG model.
To determine the nature of the IR fixed point, in this Section we
exploit a mapping onto an integrable deformation of the Z$_4$ parafermion CFT,
discovered some time ago by Fateev and Zamolodchikov \cite{fateevzam}.
The Z$_N$ parafermion CFT \cite{zamolo}  has central charge $c_N = 2 (N-1)/(N+2)$ and,
in addition to the chiral components of the stress-energy tensor,
is characterized by a chiral algebra containing
$N-1$ left (respectively, right) parafermionic currents
$\Psi_{k L}, k=1,..,N-1$ (respectively, $\Psi_{k R}$). At $N>2$ these
currents generalize the Majorana
fermion of the Ising (Z$_2$) model and are primary
fields of the Virasoro algebra with fractional spin
$h_k = k(N-k)/N$. Two integrable deformations of the
Z$_N$ CFT are known \cite{fateevzam}:
\begin{eqnarray}
{\cal H}_1 &=& {\cal H}^{0}\left(Z_N\right)
+ \lambda \left( \Psi_{1L} \Psi_{1R}
+ \Psi_{1R}^{\dagger} \Psi_{1L}^{\dagger} \right), \nonumber \\
{\cal H}_2 &=& {\cal H}^{0}\left(Z_N\right)
+ \lambda \left( e^{i\pi/N}
\Psi_{1L} \Psi_{1R}
+ e^{-i\pi/N} \Psi_{1R}^{\dagger} \Psi_{1L}^{\dagger} \right) ,
\label{intdeform}
\end{eqnarray}
where $\Psi_{1 (L,R)}^{\dagger} = \Psi_{N-1 (L,R)}$
and ${\cal H}^{0}\left(Z_N\right)$
stands for the Hamiltonian density of the Z$_N$ CFT
which has no explicit form in the general $N$ case.
Both quantum field theories have been 
studied in details and several exact results have been derived 
\cite{baseilhac,fateevnorm}.
In what follows, we shall
adopt
the prescription
that the left and right parafermionic fields commute between themselves.
\medskip

It has been shown in Ref. \cite{fateev} that the first
model in Eq. (\ref{intdeform}) with $N=4$ is equivalent to the $\beta^2 = 6\pi$
sine-Gordon model, which is a massive field theory.
The identification follows from an observation
that the two theories share the same S-matrix. On the other hand,
the second model in Eq. (\ref{intdeform}) describes an integrable
deformation of the Z$_N$ CFT characterized by a
massless flow from the UV Z$_N$ fixed point to the IR one
corresponding to
the minimal model series ${\cal M}_{N+1}$ with central
charge $1 - 6/(N+1)(N+2)$. A massless thermodynamic Bethe ansatz (TBA) system
associated with this flow has
been conjectured in Ref. \cite{fateevzam}, which in
the large-N limit reduces to that of
the O(3) non-linear sigma model with a topological term $\theta = \pi$.
For $N=4$, the massless TBA equations correspond to the $D_4$ Dynkin graph
with driving terms at two legs of the graph connected via the respective
incidence matrix.
Using the appropriate dilogarithm identities,
we have verified that the above TBA system interpolates
between $c_{UV}=1$ and $c_{IR}=4/5$, as stated in Ref. \cite{fateevzam}.
Thus, the Hamiltonian ${\cal H}_2$ with $N=4$
displays the IR critical properties falling into the
three-state Potts universality class (${\cal M}_{5}$).
It is then tempting to conjecture that
the Hamiltonian ${\cal H}_2$ with $N=4$ should be related to that of the
$\beta^2 = 6\pi$ SDSG model. In the rest of this Section we
show that they are indeed equivalent.
\medskip

It proves instructive to bosonize the two models in Eq. (\ref{intdeform}).
For arbitrary $N$, the parafermionic currents
can be expressed in terms of a suitable set of Bose fields using the Feigin-Fuchs or
coset constructions \cite{nemes,griffin,dunne}. However, the case $N=4$ is special
since the Z$_4$ theory is characterized by central
charge $c=1$ and so it should be possible to realized it by a single free
Bose field. In fact,
the identification on the level of the partition function has
been done by Yang \cite{yang}, and the Z$_4$ parafermion CFT 
with a diagonal modular invariant partition function turns out to be
equivalent to a Bose field living on the orbifold line
at radius $R=\sqrt{3/2\pi}$ in our normalization.
However, it is still possible to bosonize
the Z$_4$ parafermionic currents $\Psi_{1,2 L,R}$ with
a simple (periodic) Bose field defined on the circle
with radius $R=\sqrt{3/2\pi}$, as argued in the Appendix.
In particular, a bosonic representation for the first left
Z$_4$-parafermionic current reads
\begin{equation}
\Psi_{1L} = \frac{1}{\sqrt{2}}
\left[
:\exp\left(i \sqrt{6\pi} \; \Phi_L \right):
+ \; e^{i \sqrt{3\pi/2} \; p_{L}}
:\exp\left(-i \sqrt{6\pi} \; \Phi_L \right): \right],
\label{z4bosleft}
\end{equation}
where the
zero mode momentum operator $p_L$
has the following discrete spectrum: $p_L = n \sqrt{2\pi/3}  + m \sqrt{6\pi}$,
$n$ and $m$ being integers as a consequence of
the compactification of
the Bose field at radius $R=\sqrt{3/2\pi}$.
A similar construction can be done in the right sector by
introducing a right Bose field $\Phi_R$. So
${\cal H}_1$ and ${\cal H}_2$ in Eq. (\ref{intdeform})
can, in turn, be expressed in terms
of the total Bose field, $\Phi = \Phi_L + \Phi_R$, and its dual,
$\Theta = \Phi_L - \Phi_R$.
For the first model we find
\begin{equation}
{\cal H}_1 = \frac{1}{2} \left[ \left(\partial_x \Phi\right)^2
+ \left(\partial_x \Theta\right)^2 \right]
+ 2 \lambda :\cos\left(\sqrt{6\pi}\; \Phi\right):,
\label{sgordon}
\end{equation}
which is nothing but the $\beta^2 = 6\pi$ sine-Gordon model,
in full agreement with the findings of Ref. \cite{fateev}.
In contrast, we obtain a generalized $\beta^2 = 6\pi$ sine-Gordon model for
${\cal H}_2$:
\begin{equation}
{\cal H}_2 = \frac{1}{2} \left[ \left(\partial_x \Phi\right)^2
+ \left(\partial_x \Theta\right)^2 \right]
+ \lambda \sqrt{2} \left[:\cos\left(\sqrt{6\pi}\; \Phi\right):
+ i\; e^{i \sqrt{3\pi/2} p_L} :\cos\left(\sqrt{6\pi}\; \Theta\right):\right].
\label{selfdualbis}
\end{equation}
The cocycle operator that enters this equation takes
the values $e^{i \sqrt{3\pi/2} p_L} = \pm 1$; this follows from
the discrete spectrum of the zero-mode momentum
operator. Moreover, one should note that this cocycle operator
anticommutes with $\cos(\sqrt{6\pi}\Theta)$,
as it can be easily seen from the mode decomposition; see Eq. (\ref{modexp})
of the Appendix. This ensures hermiticity of the
Hamiltonian (\ref{selfdualbis}).
The cocycle operator in Eq. (\ref{selfdualbis})
can be absorbed in a redefinition of the dual Bose field.
Therefore, as far as the {\sl bulk} properties of the model
are concerned, the Hamiltonian (\ref{selfdualbis}) is equivalent
to that of the $\beta^2 = 6\pi$ SDSG model (\ref{actionsdsg}).
From this correspondence and from the existence of the massless flow
of the integrable deformation of the Z$_4$ CFT (${\cal H}_2$ 
in Eq. (\ref{intdeform})),
we finally conclude that the $\beta^2 = 6\pi$ SDSG model
displays the three-state Potts criticality.

%%%%%%%%%%%%%%%%%%%%%%%%%%%%%%%%%%%%%%%%%%%%%%%%%%%%%%%%%%%%%%%%%%%%%%%%%%%%%
\section{The $\beta^2 = 6 \pi$ SDSG model
and chirally stabilized liquids}

In this Section, the nature of the critical point of
the $\beta^2 = 6 \pi$ SDSG model will be
studied using a different approach.
Namely, we will make contact with a weakly perturbed,
chirally asymmetric WZNW model. Field theories of this
kind have recently attracted much interest
in connection with the universality class of chirally stabilized liquids,
introduced by Andrei, Douglas, and Jerez
\cite{andrei}.

%%%%%%%%%%%%%%%%%%%%%%%%%%%%%%%%%%%%%%%%%%%%%%%%%%%%%%%%%%%%%%%%%%%%%%%%%%%%%
\subsection{Chirally stabilized liquids}

Consider an [su(2)$_N$]$_R$ $\otimes$ [su(2)$_k$]$_L$ invariant WZNW model 
perturbed by a marginal
current-current interaction. The condition $N > k$ makes the model chirally
asymmetric. The Hamiltonian density reads:
\begin{equation}
{\cal H} = \frac{2\pi v}{N + 2} \;{\bf J}_R.{\bf J}_R
+ \frac{2 \pi v}{k + 2} \;{\bf J}_L.{\bf J}_L
+ \; g \;{\bf J}_R. {\bf J}_L,
\label{hchiral}
\end{equation}
where ${\bf J}_{R}$ and ${\bf J}_{L}$ are the right su(2)$_N$
and left su(2)$_k$ chiral vector currents, $v$ being the velocity.
The OPE for the su(2)$_N$ current in the right sector is conventionally
defined as
\begin{equation}
J_R^{a} \left(\bar z\right) J_R^{b} \left(\bar w\right) \sim
\frac{N \delta^{ab}}{8\pi^2\left(\bar z - \bar w\right)^2}
+ \frac{i \epsilon^{a b c}}{2\pi\left(\bar z - \bar w\right)}
J_R^{c} \left(\bar w\right),
\label{su2kope}
\end{equation}
with an analogous expression for the holomorphic (left)
current ($N$ to be replaced by $k$).
A simple one-loop RG analysis shows that
the interaction in Eq. (\ref{hchiral}) is marginally relevant
for a positive coupling constant $g$, so that the UV 
fixed point is unstable. When the chiral symmetry is restored ($N = k$),
the Hamiltonian
(\ref{hchiral}) is nothing but the standard su(2)$_N$
WZNW model with a marginally relevant current-current interaction.
In that case, it is well known
that this model has a spectral gap generated by the interaction.
The low energy excitations consist of
massive kinks and antikinks \cite{lowenstein,ahn}.
Since the one-loop beta function does not depend on the levels
of the su(2) KM algebras in the right and left sectors,
one might conclude that at $N>k$
the chiral asymmetry does not play any role, and in the IR limit the
system will enter a massive phase.
However, this naive picture is not correct. Instead
there is a massless flow towards an
conformally invariant fixed point.
The existence of this non-trivial criticality has been
discovered by Polyakov and Wiegmann \cite{polyakov} who
studied a related fermionic model in a special
limit where $N$ and $k$ are sent to infinity at a fixed
difference $N - k > 0$. In particular, they argued
that the criticality results from the chiral excess of particles
in the problem and  belongs to the su(2)$_{N - k}$ WZNW
universality class.
Physically, this means that a finite excess of the right movers over
the left movers makes it
impossible to bind all chiral particles into a gapped state,
so that, in the IR limit, some degrees of freedom
ought to remain massless.
\medskip

For finite values of  $N$ and $k$,
the symmetry of the IR criticality in the model (\ref{hchiral})
was shown to be \cite{andrei}
\begin{equation}
\left[su\left(2\right)_{N - k}\right]_R \otimes
\left[\frac{su\left(2\right)_k
\times su\left(2\right)_{N - k}}{su\left(2\right)_{N}}\right]_L .
\label{symmchiralfp}
\end{equation}
This result is consistent with several facts.
First of all, the global SU(2) symmetry should remain intact at the IR fixed
point. The difference between the right and left central
charges, as well as the difference between the levels of the
chiral KM algebras, should be preserved under the
RG flow. The model
is integrable and the IR central charge can be extracted
by means of the TBA approach. This has been done in Ref. \cite{andrei} and
it was found that $c_{IR}$ coincides with that of the CFT given
by Eq. (\ref{symmchiralfp}). The fixed point with the symmetry
(\ref{symmchiralfp}) has been checked for the two channel case ($N = 2, k =1$)
by utilizing the existence of a decoupling (Toulouse)
point where the mapping onto Majorana fermions solves the problem \cite{aln,azaria}.
Finally, Leclair \cite{leclair} has recently obtained
the all-orders beta-function for the model (\ref{hchiral})
confirming the existence of the stable IR fixed point at $N \ne k$.
\medskip

It is worth mentioning here that the fixed point with the symmetry (\ref{symmchiralfp})
provides
an example of a non-Fermi-liquid which is characterized by {\sl universal} critical
exponents and in this respect differs from the standard Luttinger
liquid \cite{voit}. This state has been dubbed the chirally stabilized liquid \cite{andrei}
since the very existence of the critical fixed point
follows from the chiral asymmetry of the model.
Several strongly correlated systems have been found  displaying
similar properties in the low-energy limit.
Chiral asymmetry can, of course, result from broken time reversal invariance.
An example is edge states in a paired sample of integer quantum
Hall systems with different filling factors \cite{andrei}.
On the other hand, it is also possible that
the field-theoretical Hamiltonian,
describing universal properties of a lattice model invariant under time reversal,
decomposes into two commuting and chirally asymmetric parts,
${\cal H} = {\cal H}_1 + {\cal H}_2$,
such that ${\cal H}_1$ and ${\cal H}_2$ are not separately time-reversal
invariant but transform to each other under $t \rightarrow -t$.
This scenario is realized in certain one-dimensional quantum systems,
such as the three-leg spin ladder with
crossings \cite{aln} and the
Kondo-Heisenberg chain \cite{aln,orignac,orignacbis,azaria,azariabis}.

\medskip

Revelant to our purposes is the four-channel
case ($N=4$, $k=1$) for which the symmetry
of the
IR fixed point is
\begin{equation}
\left[su\left(2\right)_3\right]_R \otimes
\left[\frac{su\left(2\right)_1
\times su\left(2\right)_3}{su\left(2\right)_4}\right]_L =
\left[su\left(2\right)_3\right]_R \otimes\left[{\cal M}_{5}\right]_L,
\label{third}
\end{equation}
where we have used, in the left sector, the coset 
construction \cite{gko} of the 
minimal model ${\cal M}_{5}$ with central charge $c=4/5$.
Expression (\ref{third})
can be further simplified using the coset \cite{zamolo}
$su(2)_3/ u(1)_3 \sim Z_3$,
where $u(1)_3$ is a rational $c=1$
CFT \cite{verlinde}  realized by a Bose field with compactification radius
$R = \sqrt{3/2\pi}$.
Finally, it is known that the Z$_3$ CFT describes
the three-state Potts model (see for instance Ref. \cite{dms}), so
that the symmetry of the IR fixed point in
the four-channel case is given by
\begin{equation}
\left[u\left(1\right)_3\right]_R \otimes Z_{3},
\label{fpfourchannel}
\end{equation}
where $Z_3$ stands for the full, chirally invariant,
three-state Potts model.
In the approach
suggested in this
Section, we first need to find a procedure
to extract those bosonic degrees of freedom from
the original Hamiltonian (\ref{hchiral}) which
account for the ``redundant'' u(1) criticality in the right sector
of Eq. (\ref{fpfourchannel}).
The
remaining degrees of freedom will then describe
the three-state Potts universality
class. The last step of
the procedure will be
to relate these degrees of freedom
to the $\beta^2 = 6 \pi$ SDSG model.

%%%%%%%%%%%%%%%%%%%%%%%%%%%%%%%%%%%%%%%%%%%%%%%%%%%%%%%%%%%%%%%%%%%%%%%%%%%%%

\subsection{Decoupling bosonic degrees of freedom}

Let us consider the original model (\ref{hchiral}) at $N=4$ and $k=1$ with
anisotropic current-current interaction:
\begin{equation}
{\cal H}
= \frac{\pi v}{3} \;{\bf J}_R.{\bf J}_R + \frac{2 \pi v}{3} \;
{\bf J}_L.{\bf J}_L + g_{\parallel} \; J_R^{z} J_L^{z}
+ \frac{g_{\perp}}{2} \left(J_R^{+} J_L^{-} + H. c. \right).
\label{h1h2brok}
\end{equation}
We will assume that $g_{\parallel} > 0$. This condition ensures that the model
flows towards strong coupling in the IR limit.
The sign of the transverse coupling constant, $g_{\perp}$,
is arbitrary ($g_{\perp} \to -  g_{\perp}$ under the transformations
$J_R^{\pm} \rightarrow - J_R^{\pm}, J_R^z \rightarrow J_R^z$).
The important feature of the anisotropic model (\ref{h1h2brok}) is that
it displays the same universal behavior (\ref{fpfourchannel}) as in the 
fully SU(2)-symmetric case.
Restoration of the SU(2) symmetry at IR fixed point can be inferred from
the recently computed all-orders beta function of 
the model (\ref{h1h2brok}) \cite{leclair}
and, since the anisotropic version is still integrable, by means of the
Bethe-ansatz approach \cite{alexeicom}. The advantage of the anisotropic model
is that it becomes particularly simple in the so-called Toulouse limit,
similar to the decoupling (or Luther-Emery \cite{luther}) point in the 
sine-Gordon
model. There exists a special line in the parameter space of
the coupling constants along which certain bosonic degrees of freedom
decouple from the rest of the spectrum and remain massless in
the IR limit. Toulouse type solutions proved extremely fruitful in recent years,
especially in quantum impurity problems \cite{bookboso} and,
most notably, in the two-channel Kondo problem \cite{emerykiv}.
\medskip

The starting point of the Toulouse solution is the introduction
of a bosonized
description for the su(2)$_4$ KM current.
Such identification can be derived from the
coset $su(2)_4/ u(1)_4 \sim Z_4$
which relates the su(2)$_4$ KM algebra
to the Z$_4$ parafermion CFT through the u(1)$_4$ rational CFT,
corresponding to a Bose field living on the circle at the
radius $R = \sqrt{2/\pi}$.
The right su(2)$_4$ current (${\bf J}_{R}$) can then be expressed
in terms of a chiral Bose field $\Phi_{sR}$
and the first Z$_4$ parafermion
current $\psi_{1R}$ via
\begin{eqnarray}
J_R^{\dagger} &=& \frac{1}{\pi}\; \psi_{1R} \;
:\exp\left(-i \sqrt{2\pi} \; \Phi_{sR}
\right): \nonumber \\
J_R^{-} &=& \frac{1}{\pi} \; :\exp\left(i \sqrt{2\pi} \; \Phi_{sR}
\right): \psi_{1R}^{\dagger} \nonumber \\
J_R^{z} &=& -\sqrt{\frac{2}{\pi}} \; i {\bar \partial} \Phi_{sR} =
\sqrt{\frac{2}{\pi}} \; \partial_x \Phi_{sR},
\label{bosu2Nleft}
\end{eqnarray}
with the prescription that
the parafermionic fields
commute with the bosonic ones.
Then, using the parafermionic algebra \cite{zamolo}
it is not difficult to show that
the correspondence (\ref{bosu2Nleft}) reproduces
OPE (\ref{su2kope}) with $N=4$.
In the left sector, we introduce a Bose field at the self-dual
radius $R_0 = 1/\sqrt{2\pi}$ to
obtain a bosonized description of the
su(2)$_1$ KM current as in Eq. (\ref{su21freerep})
\begin{eqnarray}
J_L^+ &=& \frac{1}{2\pi} \;:\exp\left( i \sqrt{8\pi}\Phi_{0L}\right):
\nonumber \\
J_L^z &=& \frac{i}{\sqrt{2\pi}} \; \partial \Phi_{0L}
      = \frac{1}{\sqrt{2\pi}} \; \partial_x \Phi_{0L}.
\label{su21freerepbis}
\end{eqnarray}
(we are working here with
left and right moving Bose fields that commute with themselves).
\medskip

The Hamiltonian (\ref{h1h2brok}) can be expressed in terms of
the bosonic and parafermionic fields as follows:
\begin{eqnarray}
{\cal H} &=& v\left[\left(\partial_x \Phi_{0L}\right)^2 +
\left(\partial_x \Phi_{sR}\right)^2 \right]
+ \frac{g_{\parallel}}{\pi} \; \partial_x \Phi_{0L}
\partial_x \Phi_{sR}
+ {\cal H}^{0}_{R}\left(Z_4\right)
\nonumber \\
&+& \frac{g_{\perp}}{4\pi^2}
\left[\psi_{1R} :\exp\left(-i\sqrt{2\pi} \Phi_{sR} - i
\sqrt{8\pi} \Phi_{0L} \right):  +
H.c. \right],
\label{h1boso}
\end{eqnarray}
where ${\cal H}^{0}_{R}\left(Z_4\right)$ is the
right-moving piece of the Hamiltonian density associated to the Z$_4$ CFT.
The cross derivative terms in Eq. (\ref{h1boso}) can be eliminated
by performing a canonical transformation of the Bose
fields (this is the standard procedure that solves the Luttinger model;
see e.g. Ref. \cite{voit}):
\begin{eqnarray}
\left(\begin{array}{c}
\Phi_{0L}\\
\Phi_{sR}
\end{array}\right) =
\left(
\begin{array}{lccr}
{\rm ch} \alpha & {\rm sh} \alpha \\
{\rm sh} \alpha & {\rm ch} \alpha
\end{array} \right) \left(\begin{array}{c}
\Phi_{2L}\\
\Phi_{1R}
\end{array}\right),
\label{can1}
\end{eqnarray}
with
\begin{equation}
{\rm th} 2 \alpha = -
\frac{g_{\parallel}}{2 \pi v}.
\label{th1}
\end{equation}
Under this transformation, the
argument of the vertex operator in Eq. (\ref{h1boso})
becomes
\begin{equation}
\sqrt{2\pi} \Phi_{sR} + \sqrt{8\pi} \Phi_{0L}
\rightarrow \sqrt{2\pi} \left[\left(
2\; {\rm ch} \alpha + {\rm sh} \alpha\right) \Phi_{2L}
+ \left({\rm ch} \alpha +  2 \; {\rm sh} \alpha \right)
\Phi_{1R}
\right].
\label{costran}
\end{equation}
We then observe that for a special value of $\alpha$  
determined by the condition
\begin{equation}
{\rm th} \alpha = -\frac{1}{2},
\label{th2}
\end{equation}
(the corresponding (nonuniversal) value of the coupling $g_{\parallel}$
is $g_{\parallel}^{*} = 8 \pi v/5$) the Bose field $\Phi_{1R}$ decouples from the
rest of the Hamiltonian, so that
the degrees of freedom described by this field will remain critical.
This is the main feature of the Toulouse point.
In the new basis, the Hamiltonian (\ref{h1boso}) transforms to
\begin{equation}
{\cal H} =
u\left[ \left(\partial_x \Phi_{1R}\right)^2
+ \left(\partial_x  {\Phi}_{2L}\right)^2 \right]
+ {\cal H}^{0}_{R}\left(Z_4\right)
+ \frac{g_{\perp}}{4\pi^2}
\left[\psi_{1R} :\exp\left(-i\sqrt{6\pi}
{\Phi}_{2L}\right):
+ H.c. \right],
\label{h1toulouse}
\end{equation}
with the  renormalized velocity $u = 3 v/5$.
At the Toulouse point,
the transformation (\ref{can1}) of the chiral Bose fields
is simplified:
\begin{eqnarray}
\Phi_{0L} &=& \frac{1}{\sqrt{3}}\left( 2 \Phi_{2L}
- \Phi_{1R} \right) \nonumber \\
\Phi_{sR} &=&
\frac{1}{\sqrt{3}}\left( 2 \Phi_{1R}
- \Phi_{2L}\right),
\label{cantou}
\end{eqnarray}
the inverse transformation being
\begin{eqnarray}
\Phi_{1R} &=& \frac{1}{\sqrt{3}}\left( 2 \Phi_{sR}
+  \Phi_{0L} \right) \nonumber \\
\Phi_{2L}  &=&
\frac{1}{\sqrt{3}}\left(2 \Phi_{0L} + \Phi_{sR}
\right).
\label{cantouinv}
\end{eqnarray}

%%%%%%%%%%%%%%%%%%%%%%%%%%%%%%%%%%%%%%%%%%%%%%%%%%%%%%%%%%%%%%%%%%%%%%%%%%%%%
\subsection{Three-state Potts criticality}

At this point, let us pause to discuss the situation at hand.
Starting from the anisotropic model (\ref{h1h2brok}),
the decoupling of the right-moving bosonic degree of freedom
described by the field $\Phi_{1R}$ has been achieved
at the Toulouse point.
Moreover, this Bose field is compactified with radius
$R = \sqrt{3/2\pi}$, as can be seen from
Eq. (\ref{cantouinv}) and the values of the radii
for $\Phi_0$ and $\Phi_s$. We thus deduce that the Bose field $\Phi_{1R}$
describes a chiral $[u(1)_3]_R$ rational CFT.
The question is how to interpret the remaining degrees of freedom
described by the Bose field $\Phi_{2L}$
and the Z$_4$ parafermionic field $\psi_{1R}$ which,
according to Eq. (\ref{h1toulouse}), are nontrivially coupled.
As already stressed in the preceeding subsection,
in the anisotropic model (\ref{h1h2brok}) the SU(2) symmetry is
asymptotically restored in the IR limit and the symmetry of 
the IR fixed point is thus still given
by Eq. (\ref{fpfourchannel}). This observation leads us to conclude
that the Hamiltonian
\begin{equation}
{\cal H}_{\rm eff} =
u \left(\partial_x{\Phi}_{2L}\right)^2
+ {\cal H}^{0}_{R}\left(Z_4\right)
+ \frac{g_{\perp}}{4\pi^2}
\left[\psi_{1R} :\exp\left(-i\sqrt{6\pi}
{\Phi}_{2L}\right):
+ H.c. \right],
\label{toueffpotts}
\end{equation}
describes critical properties of the three-state Potts model.
\medskip

Thus, we are left to relate the Hamiltonian (\ref{toueffpotts})
to the $\beta^2 = 6 \pi$ SDSG model.
To this end, we introduce a right-moving Bose field
$\varphi_R$ to bosonize the Z$_4$ parafermion fields which
enter Eq. (\ref{toueffpotts}).
Using representation (\ref{z4currbos3}) of the Appendix in
the right sector, we transform the
Hamiltonian (\ref{toueffpotts}) to
\begin{eqnarray}
{\cal H}_{\rm eff} = u\left[\left(\partial_x \Phi_{2L}\right)^2
+ \left(\partial_x \varphi_{R}\right)^2\right]
+ \frac{g_{\perp}\sqrt{2}}{4\pi^2}\left[
:\cos\left(\sqrt{6\pi}\left(\varphi_R - \Phi_{2L}\right)
\right):  \right. \nonumber \\
\left.
- i \; e^{-i \sqrt{3 \pi/2} p_R}
:\sin\left(\sqrt{6\pi}\left(\varphi_R + \Phi_{2L}\right)
\right):\right],
\label{hfourchanbos}
\end{eqnarray}
where we have neglected the velocity anisotropy between
the two Bose fields.
The zero mode momentum $p_R$ associated
with the chiral Bose field $\varphi_R$
has a discrete spectrum: $p_R = n\sqrt{2\pi/3} - m\sqrt{6\pi}$,
so that the
cocycle operator in Eq. (\ref{hfourchanbos})
takes the values: $e^{i \sqrt{3 \pi/2} \;p_{R}} = \pm 1$.
The two chiral Bose fields in Eq. (\ref{hfourchanbos})
can be combined into a total
Bose field $\Phi = \varphi_R + \Phi_{2L}$ and its dual
field $\Theta = \Phi_{2L} - \varphi_R$.
The above Hamiltonian
then simplifies as
\begin{equation}
{\cal H}_{\rm eff} = \frac{u}{2}\left[\left(\partial_x \Phi\right)^2
+ \left(\partial_x \Theta\right)^2\right]
+ \frac{g_{\perp}\sqrt{2}}{4\pi^2}\left[
:\cos\left(\sqrt{6\pi} \Theta
\right):
- i \; e^{-i \sqrt{3 \pi/2} p_R}
:\sin\left(\sqrt{6\pi}\Phi
\right):\right].
\label{selfdual}
\end{equation}
By absorbing the cocycle operator into the Bose field,
the resulting Hamiltonian shares the same bulk properties as
\begin{equation}
{\cal H}_{\rm eff} = \frac{u}{2}\left[\left(\partial_x \Phi\right)^2
+ \left(\partial_x \Theta\right)^2\right]
+ \frac{g_{\perp}\sqrt{2}}{4\pi^2}\left[
:\cos\left(\sqrt{6\pi} \Theta
\right):
+
:\cos\left(\sqrt{6\pi}\Phi
\right):\right],
\label{selfdualham}
\end{equation}
which is nothing but the Hamiltonian of the $\beta^2 = 6 \pi$ SDSG model.
Thus we arrive at the same conclusion as in the end of section III
i.e. the $\beta^2 = 6 \pi$ SDSG model displays critical properties
in the three-state Potts universality class.

%%%%%%%%%%%%%%%%%%%%%%%%%%%%%%%%%%%%%%%%%%%%%%%%%%%%%%%%%%%%%%%%%%%%%%%%%%%%%
\subsection{The UV-IR transmutation of the fields}

The next important point is the determination 
of the UV-IR transmutation associated with the 
massless flow of the $\beta^2 = 6 \pi$ SDSG model. 
To this end, one has to find out how the operators,
originally defined in the vicinity of the UV fixed point,
transmute into the three-state Potts fields when going from 
the UV limit to the IR limit.
Unfortunately, this is not an easy task since 
the $\beta^2 = 6 \pi$ SDSG model is a non-trivial 
field theory. In particular, it does not admit any simple 
decomposition between the massive and critical degrees of freedom
which was crucial a step to perform the UV-IR transmutation 
of the DSG model \cite{fgn}.
Futhermore, the UV-IR correspondence of 
the integrable deformation of the Z$_4$ parafermion
theory (\ref{intdeform}), equivalent to the $\beta^2 = 6 \pi$ SDSG model,
is unknown to the best of our knowledge.
However, we shall here present some {\it conjectures}
on the UV-IR transmutation of some fields using 
the massless flow of the chirally asymmetric su(2)$_4$ $\otimes$
su(2)$_1$ WZNW model (\ref{hchiral}) and the Toulouse limit solution.

Let us first discuss more precisely the UV limit of the 
$\beta^2 = 6 \pi$ SDSG model (\ref{selfdualham}).
It corresponds to a $c=1$ CFT described by a bosonic 
field living on the circle at radius $R = \sqrt{3/2\pi}$.
At this special radius, the resulting u(1)$_3$ CFT 
exhibits an extended symmetry algebra ${\cal A}_3$ \cite{verlinde}
which is generated in the left sector by the standard left 
u(1) current $J_L = i \partial \Phi_L$ together with extra 
left  currents with spin 3: 
$\Gamma^{\pm}_L = :\exp\left(\pm i \sqrt{24\pi} \Phi_L\right):$.
Under this extended algebra,
the u(1)$_3$ CFT  has
a finite number of primary fields and is thus an
example of a rational CFT.
The partition function of this CFT on the torus is given
by Eq. (\ref{partbos}) of the Appendix and the 
six primary fields are 
vertex operators $V_{\lambda}$ mutually local 
with the currents of the extended symmetry:
\begin{equation}
V_{\lambda} = :\exp\left(i \lambda \sqrt{2\pi/3} \; \Phi\right):,
\label{primaryu13}
\end{equation}
with $\lambda = 0, \pm 1, \pm 2, 3$ and conformal weights
$(\lambda^2/12, \lambda^2/12)$.
Interestingly, one can associate a Z$_3$ charge $q$ 
to the vertex operators $V_{\lambda}$ (\ref{primaryu13}) through:  
$q \equiv \lambda \; ({\rm mod} \; 3)$, $\lambda$ being integer.
This charge is additive under fusion of these 
vertex operators. This Z$_3$ symmetry is simply generated 
by a shift on the Bose field: 
$\Phi \rightarrow \Phi + \sqrt{2\pi/3}$. We 
note that the perturbing field of 
the $\beta^2 = 6 \pi$ SDSG model is neutral with respect to this 
Z$_3$ symmetry.
In fact, the explicit description of the Z$_3$ symmetry
of the $\beta^2 = 6 \pi$ SDSG model can also be 
derived by considering its lattice version i.e. the two-dimensional 
classical XY model with a three-fold symmetry breaking
field (see Eq. (\ref{xyanisoham}) with $N=3$).
The Z$_3$ symmetry of this model is described by
the following transformation on the lattice variable $\theta_{\bf r}$: 
$\theta_{\bf r} \rightarrow \theta_{\bf r} + 2 \pi/3$  so that
in the continuum limit it will correspond to
$\Phi \rightarrow \Phi + \sqrt{2\pi/3}$ as can be easily seen.

We now turn to the analysis of the UV-IR correspondence
of the $\beta^2 = 6 \pi$ SDSG model by considering
the chirally asymmetric su(2)$_4$ $\otimes$
su(2)$_1$ WZNW model (\ref{hchiral}).
As already discussed, this latter model is characterized 
by the following massless flow:
\begin{equation}
\left[su\left(2\right)_1\right]_L \otimes
\left[su\left(2\right)_4\right]_R \rightarrow
\left[su\left(2\right)_3\right]_R \otimes
\left[\frac{su\left(2\right)_1
\times su\left(2\right)_3}{su\left(2\right)_4}\right]_L \sim
\left[u\left(1\right)_3\right]_R \otimes Z_3 .
\label{massflowchir}
\end{equation}
The resulting UV-IR transmutation can be analysed 
from the conservation of the su(2) spin and one has
the following correspondence \cite{andrei,orignacbis}: 
\begin{equation}
\left[\Phi^{(l_1)}_1\right]_L  
\left[\Phi^{(l_2)}_4\right]_R \sim 
\sum_{|l_1 - l_2| \le l_3 \le l_1 + l_2}
\left[\Phi^{(l_3)}_3\right]_R \varphi^{l_1 l_3}_{l_2 L},
\label{transchir}
\end{equation}
where the su(2)$_k$ primaries are denoted by $\Phi^{(l)}_k$
with $l=0,1,..,k$ and carry spin $l/2$.
In Eq. (\ref{transchir}), $\varphi^{l_1 l_3}_{l_2 L}$
($0 \le l_1 \le 1, 0 \le l_2 \le 4, 0 \le l_3 \le 3$)
are the left fields of the coset in Eq. (\ref{massflowchir})
with conformal weight $l_1(l_1+2)/12 + l_3(l_3+2)/20 - l_2(l_2+2)/24$.
Branching selection rules restrict $l_1 + l_2 + l_3$
to be even and one has also the field identification:
$\varphi^{l_1 l_3}_{l_2 L} \sim \varphi^{1- l_1 3 - l_3}_{4 - l_2 L}$.

The strategy  to determine the UV-IR correspondence 
of the $\beta^2 = 6 \pi$ SDSG model is to choose 
special values of $l_1$ and $l_2$ so that the primary fields
in the lhs of Eq. (\ref{transchir}) have a simple free-field representation. 
The next step is to make use of the Toulouse
basis (\ref{cantou}) to express these fields in terms 
of the $\beta^2 = 6 \pi$ SDSG UV operators and vertex
operators built from the $\Phi_{1R}$ bosonic field.
By extracting the IR u(1) criticality
in Eq. (\ref{massflowchir}) associated with the field $\Phi_{1R}$, 
one can expect to obtain a representation of some IR operators
of the three-state Potts model in terms of the original 
UV fields of the $\beta^2 = 6 \pi$ SDSG model.
For instance, the current-current perturbation 
of the su(2)$_4$ $\otimes$ su(2)$_1$ WZNW model (\ref{hchiral}) 
is known  \cite{orignac,orignacbis}
to have conformal 
weights $(7/5,7/5)$ in the IR limit and thus
identifies to the neutral X field of the three-state Potts model.
As described in the Toulouse limit approach of this model, 
the UV perturbation reduces to the self-dual contribution of 
Eq. (\ref{selfdualham}) so that one expects 
the following UV-IR correspondence:
\begin{equation}
:\cos\left(\sqrt{6\pi} \Phi 
\right): + :\cos\left(\sqrt{6\pi}\Theta \right): \;
\sim X .
\label{firstrans}
\end{equation}
One should note that this result is consistent with the fact
that these two operators are neutral with respect to the 
Z$_3$ symmetry and the charge conjugation.
In fact, the transmutation (\ref{firstrans}) can also
be justified by an additional argument.
Indeed, it is known \cite{fateevzam} that the perturbation of ${\cal H}_2$
in Eq. (\ref{intdeform}) degenerates, in the IR limit, 
into the irrelevant operator $\Phi_{31}$ of the ${\cal M}_{N+1}$ 
minimal model with  
conformal weights $((N+3)/N+1, (N+3)/N+1)$.
In the special $N=4$ case, the perturbation 
of ${\cal H}_2$  is nothing but the SDSG perturbing field
as seen in Section III so that one recovers the 
correspondence (\ref{firstrans}).

Further progress can be made by considering 
Eq. (\ref{transchir})  with $l_1 =0$ and $l_2= l_3 = 2$.
In that case, the su(2)$_4$ primary field $\Phi^{(2)}_4$, transforming
in the spin 1 representation, has a free-field description \cite{ZF} in terms 
of two Bose fields using the conformal embedding 
su(2)$_4$ $\in$ su(3)$_1$. 
On the other hand, the su(2)$_3$ primary 
field $\Phi^{(2)}_3$ can be expressed
in terms of the Z$_3$ spin field and  the chiral bose field $\Phi_{1R}$
using Eq. (\ref{suparaprim}).
With help of the Toulouse basis (\ref{cantou}),
we have obtained that the leading behavior in the
IR limit of the vertex operators 
$V_{\pm 1}$ (\ref{primaryu13}), which carry a $q= \pm 1$ Z$_3$ 
charge as discussed above, identifies to the 
two Z$_3$ spin fields $\sigma$ and $\sigma^{\dagger}$ with
conformal weights $(1/15,1/15)$ and $q= \pm 1$ Z$_3$ charge:
\begin{eqnarray}
:\exp\left(i \sqrt{2\pi/3} \; \Phi\right): \; &\sim&
\; \sigma  \nonumber \\
:\exp\left(-i \sqrt{2\pi/3} \; \Phi\right): \; &\sim&
\; \sigma^{\dagger} .
\label{secondtrans}
\end{eqnarray}
A similar approach can be applied for $l_1 = 1$ and $l_2 = 2$ 
where now two terms in the rhs of Eq. (\ref{transchir}) contribute
with $l_3 = 1$ and $l_3 = 3$.
As for the left su(2)$_1$ current (\ref{su21freerepbis}),
the left su(2)$_1$ primary field $\Phi^{(1)}_1|_L$ has a simple free-field
representation in terms of the Bose field $\Phi_{0L}$ \cite{ZF}.
In this case, we have obtained the following 
UV-IR correspondence:
\begin{eqnarray}
:\exp\left(- i \; 2 \sqrt{2\pi/3} \; \Phi\right): \; &\sim&
\; \sigma  + \psi_{1R} \psi_{1L} \nonumber \\
:\exp\left(i \; 2 \sqrt{2\pi/3} \; \Phi\right): \; &\sim&
\; \sigma^{\dagger} + \psi_{1R}^{\dagger} \psi_{1L}^{\dagger},
\label{thirdtrans}
\end{eqnarray}
where $\psi_{1R}$ (respectively $\psi_{1L}$) is the 
right (respectively left) Z$_3$ parafermionic current 
with conformal weights $(0,2/3)$
(respectively $(2/3,0)$).  The primary field $\psi_{1R} \psi_{1L}$
with conformal weights $(2/3,2/3)$, also denoted by Z$_1$ in 
the book \cite{dms}, has a $q=1$ Z$_3$ charge.
We observe that the result (\ref{thirdtrans})
is consistent with 
the Z$_3$ charge of the vertex operators (\ref{primaryu13})
through the transformation $\Phi \rightarrow \Phi  + \sqrt{2\pi/3}$.
%%%%%%%%%%%%%%%%%%%%%%%%%%%%%%%%%%%%%%%%%%%%%%%%%%%%%%%%%%%%%%%%%%%%%%%%%%%%

\section{Concluding remarks}

In this paper, we have discussed the critical properties
of the $\beta^2 = 2\pi N$ SDSG model
which provides a continuum description of the two-dimensional classical
XY model with an N-fold symmetry breaking field.
This system  has a
single phase transition for $N=2$ and $N=3$
which falls into the Ising and three-state Potts
universality class, respectively.
The $N=4$ case exhibits continuously varying critical
exponents typical for the Luttinger-liquid behavior.
The $N=2$ and $N=4$ criticalities can be
clearly understood and described starting from
the corresponding SDSG model and treating it by standard methods,
like bosonization or perturbative RG approaches. The case $N=3$
is exceptional for being strongly non-perturbative
and resistant to any simple-minded free-field treatment.

The three-state Potts universality class of the
IR fixed point of the $\beta^2 = 6\pi$ SDSG model has been
determined in this paper by two independent approaches.
We have first mapped this model onto an integrable deformation
of the Z$_4$ parafermion CFT \cite{fateevzam}, which has a massless
flow to a three-state Potts IR fixed point.
The second approach was based on establishing a
relationship between the $\beta^2 = 6\pi$ SDSG model and
an anisotropic, chirally asymmetric version of the WZNW model
with su(2)$_4$ and su(2)$_1$ current-current interaction.
This model exhibits critical
properties of the chirally stabilized liquid universality
class \cite{andrei}.
From the nature of the IR fixed point of the latter model
we have deduced the same Z$_3$ IR properties
of the $\beta^2 = 6\pi$ SDSG model.
\medskip

Regarding perspectives,
the $\beta^2 = 6\pi$ SDSG model may be analyzed
using the form factor perturbation theory \cite{delfinoetal}
with help of the form factors of topologically
charged operators in the sine-Gordon model \cite{lukyanov}.
It will be very interesting to
determine the complete UV-IR transmutation of the
fields of the $\beta^2 = 6\pi$ SDSG model 
and in particular to check the conjectures presented in this paper.
Finally, there are specific physical
realizations of the $\beta^2 = 6\pi$ SDSG model
in one-dimensional quantum spin/electron systems
and in two-dimensional statistical mechanics.
In this respect, Delfino \cite{delfino} has recently
proposed that the $\beta^2 = 6\pi$ SDSG model
describes the field theory corresponding to the crossover
from antiferromagnetic to ferromagnetic three-state Potts
behavior.
The connection between the $\beta^2 = 6\pi$ SDSG model
and the chirally stabilized liquids, discussed in this paper, leads us to
a conclusion that this model also accounts for the Z$_3$ critical properties
of the four-channel underscreened Kondo-Heisenberg chain with incommensurate
fillings. We hope that other applications of the
$\beta^2 = 6\pi$ SDSG model will be found in the future.

%%%%%%%%%%%%%%%%%%%%%%%%%%%%%%%%%%%%%%%%%%%%%%%%%%%%%%%%%%%%%%%%%%%%%%%%%%%%%
\begin{acknowledgments}

The authors are grateful to P. Azaria for collaboration at
the initial stages of this work.
We would like also to thank D. Allen, E. Orignac,
and A. M. Tsvelik for valuable discussions.
A.O.G.'s research is supported by the EPSRC of the UK 
under grant GR/N19359 and GR/R70309.
One of us (P. L.) would like to dedicate this 
paper to his grandfather.
\end{acknowledgments}

%%%%%%%%%%%%%%%%%%%%%%%%%%%%%%%%%%%%%%%%%%%%%%%%%%%%%%%%%%%%%%%%%%%%%%%%%%%%%
\appendix
\section{Bosonization of the Z$_4$ parafermion theory}

In this Appendix, a bosonization approach
to the Z$_4$ parafermion
CFT is presented. This theory has central charge $c=1$, which
suggests that it can be brought to correspondence with
a suitably defined free Bose field.
The precise identification requires full knowledge of the
operator content of the Z$_4$ parafermion CFT.

%%%%%%%%%%%%%%%%%%%%%%%%%%%%%%%%%%%%%%%%%%%%%%%%%%%%%%%%%%%%%%%%%%%%%%%%%%%%%
\subsection{Identification of the Bose field}

The spectrum of the Z$_N$ parafermion theory
can be obtained from the $su(2)_N/u(1)_N$ coset model.
In the holomorphic sector, the su(2)$_N$ primaries ($\Phi_{m}^{l}$)
are related to the Z$_N$ parafermionic ones ($f_{m}^{l}$)
by \cite{zamolo,gepner}:
\begin{equation}
\Phi_{m}^{l} = f_{m}^{l} :\exp\left(i m \sqrt{\frac{2\pi}{N}}
\; \Phi_L\right):,
\label{suparaprim}
\end{equation}
where $l=0,..,N$ and $-N +1 \le m \le N$ with the 
constraint: $l \equiv m \; (2)$.
The operator content of the Z$_N$ CFT is obtained
by constructing different modular invariants of this series
which can be determined with help of
the coset $su(2)_N/u(1)_N$ \cite{gepner}:
\begin{equation}
{\cal Z}\left(Z_N\right) = \frac{|\eta|^2}{2} \sum_{l,{\bar l} = 0}^{N}
\sum_{m,{\bar m} = -N +1}^{N} L_{l, {\bar l}} M_{m, {\bar m}}
c_{m}^{l} c_{{\bar m}}^{{\bar l} *}
\frac{\left(1 + \left(-1\right)^{l-m}\right)
\left(1 + \left(-1\right)^{{\bar l}-{\bar m}}\right)}{4},
\label{partparagen}
\end{equation}
where $\eta$ is the Dedekind function: $\eta(q) = q^{1/24}
\prod_{n=1}^{+\infty} (1 - q^n)$. The coefficients
$c_{m}^{l}$ are the so-called level-N string functions
of the current algebra (see for instance Ref. \cite{dms})
and verify the following properties:
$c_{m}^{l} = c_{-m}^{l}$, $c_{m}^{l} = c_{N-m}^{N-l}$.
In Eq. (\ref{partparagen}), $ L_{l, {\bar l}}$ and $M_{m, {\bar m}}$
are two positive integers which define different
modular invariants of the su(2)$_N$ and u(1)$_N$ theories, respectively.
The simplest modular invariant of the Z$_4$ CFT is the diagonal
one with $ L_{l, {\bar l}} = \delta_{l, {\bar l}}$ and
$ M_{m, {\bar m}} = \delta_{m, {\bar m}}$ 
so that the resulting partition function reads:
\begin{equation}
{\cal Z}_{diag}\left(Z_4\right) = |\eta|^2 \left(
|c_{0}^{0}|^2 + 2 |c_{2}^{0}|^2 + |c_{4}^{0}|^2 +
2 |c_{1}^{1}|^2 + 2 |c_{3}^{1}|^2 + |c_{0}^{2}|^2 +
|c_{2}^{2}|^2 \right).
\label{4partparadiag}
\end{equation}
At the next step we use the
following identities for the string functions first obtained
by Yang \cite{yang}:
\begin{eqnarray}
\eta c_{0}^{0} + \eta c_{4}^{0} &=& K_{0}^{(6)} \nonumber \\
\eta c_{0}^{0} - \eta c_{4}^{0} &=& \frac{1}{\eta} \sum_{n=-\infty}^{+\infty}
\left(-1\right)^n q^{n^2} \nonumber \\
\eta c_{2}^{2} &=& K_{1}^{(6)} \nonumber \\
\eta c_{0}^{2} &=& K_{2}^{(6)} \nonumber \\
2\eta c_{2}^{0} &=& K_{3}^{(6)} \nonumber \\
\eta c_{1}^{1} &=& K_{1}^{(8)} \nonumber \\
\eta c_{3}^{1} &=& K_{3}^{(8)}.
\label{stringiden}
\end{eqnarray}
Here
\begin{equation}
K_{\lambda}^{(N)} = \frac{1}{\eta} \sum_{n=-\infty}^{+\infty}
q^{N/2\left(n + \lambda/N\right)^2},
\end{equation}
represent the generalized characters of a Bose field
living on a circle at a rational radius \cite{verlinde,dms}.
The partition function (\ref{4partparadiag}) takes then the form
\begin{eqnarray}
{\cal Z}_{diag}\left(Z_4\right) &=& \sum_{\lambda=1}^{2} |K_{\lambda}^{(6)}|^2
+ 2 |\frac{K_{3}^{(6)}}{2}|^2 + 2 \sum_{\lambda=1,3} |K_{\lambda}^{(8)}|^2
\nonumber \\
&+& |\frac{1}{2\eta} \sum_{n=-\infty}^{+\infty} q^{3n^2} +
\frac{1}{2\eta} \sum_{n=-\infty}^{+\infty} \left(-1\right)^n q^{n^2}|^2
\nonumber \\
&+& |\frac{1}{2\eta} \sum_{n=-\infty}^{+\infty} q^{3n^2} -
\frac{1}{2\eta} \sum_{n=-\infty}^{+\infty} \left(-1\right)^n q^{n^2}|^2 .
\label{paradiagorb}
\end{eqnarray}
This expression is nothing but the partition function
of a Bose field on the orbifold line at the
special radius $R=\sqrt{3/2\pi}$ (see for instance 
Ref. \cite{dms}). 
The twisted sector
of the  orbifold model corresponds to
the states with conformal weights $(h, {\bar h}) =
(1/16 + n, 1/16 + n)$ or $(9/16 + n, 9/16 + n)$ and thus
identifies with the third term in Eq. (\ref{paradiagorb}).
From the identification (\ref{stringiden}), one then deduces
that the Z$_4$ parafermion fields characterized by an odd-integer $l$
belong to the twisted sector of the orbifold theory. In contrast,
the fields with even $l$ have representations in the untwisted sector
and thus can be described by a free boson living on a circle of
radius $R = \sqrt{3/2\pi}$.
\medskip

In fact, this result can be better seen by considering
another modular invariant of the Z$_4$ CFT. Indeed,
one can replace in the general partition function (\ref{partparagen})
the diagonal su(2)$_4$ modular invariant by the non-diagonal one
which is diagonal under a larger algebra (su(3)$_1$).
In that case, the new partition function reads
\begin{equation}
{\cal Z}^{'}\left(Z_4\right) = |\eta|^2 \left(
|c_{0}^{0}+ c_{4}^{0}|^2 + 4 |c_{2}^{0}|^2 +
2 |c_{0}^{2}|^2 +
2 |c_{2}^{2}|^2 \right).
\label{4partparanondiag}
\end{equation}
We thus observe that only the parafermionic fields with even $l$
appear in this modular invariant and all the states with odd $l$
have been projected out. Using identities (\ref{stringiden}),
the partition function (\ref{4partparanondiag}) can then
be expressed in terms of the characters of the bosonic theory:
\begin{eqnarray}
{\cal Z}^{'}\left(Z_4\right) &=& |K_{0}^{(6)}|^2  + 2 |K_{1}^{(6)}|^2
+ 2 |K_{2}^{(6)}|^2 + |K_{3}^{(6)}|^2 \nonumber \\
&=& \sum_{\lambda=0}^{5} |K_{\lambda}^{(6)}|^2,
\label{partbos}
\end{eqnarray}
which is precisely the partition function of a Bose field
living on the circle at radius $R=\sqrt{3/2\pi}$.
In this paper, we are only concerned with
the bosonization of some Z$_4$ parafermionic fields with
even $l$ so that one can safely work with a free boson
on the circle with radius $R=\sqrt{3/2\pi}$.

%%%%%%%%%%%%%%%%%%%%%%%%%%%%%%%%%%%%%%%%%%%%%%%%%%%%%%%%%%%%%%%%%%%%%%%%%%%%%

\subsection{Bosonization of the Z$_4$ parafermionic currents}

We give now a bosonized description
of the Z$_4$ parafermionic currents
$\psi_{1L},~ \psi_{2L} =  \psi_{2L}^{\dagger}$, and $\psi_{1L}^{\dagger} =
\psi_{3L}$, which act on the left sector and have
dimensions $3/4, 1, 3/4$, respectively. These fields appear in the
parafermionic modules characterized by $l=0$ or $l=4$, so that it is
possible to find their bosonic representation in terms of
a compactified boson at radius $R=\sqrt{3/2\pi}$, as 
discussed above.
An explicit realization of the algebra
of the Z$_4$ parafermionic currents \cite{zamolo}         
in terms of a Bose field and some
appropriate cocycles can be derived. The latter degrees of freedom are of
utmost importance to achieve a faithful representation
of the algebra. One way  to formally implement the cocycles is to
introduce extra degrees of freedom like, for instance, the Pauli matrices
$\sigma_{x,y,z}$.
In this respect, we have checked that 
a faithfull representation of 
the Z$_4$ parafermionic currents in terms of
a chiral bosonic field $\varphi_L$ is given by 
\begin{eqnarray}
\psi_{1L} &=& \frac{e^{i \pi/4}}{\sqrt{2}}
\left( \sigma_y :\exp\left(i \sqrt{6\pi} \; \varphi_L \right):
+ \; \sigma_x :\exp\left(-i \sqrt{6\pi} \; \varphi_L \right): \right)
\nonumber \\
\psi_{1L}^{\dagger} &=& \frac{e^{-i \pi/4}}{\sqrt{2}}
\left( \sigma_y :\exp\left(-i \sqrt{6\pi} \; \varphi_L \right):
+ \; \sigma_x :\exp\left(i \sqrt{6\pi} \; \varphi_L \right): \right)
\nonumber \\
\psi_{2L} &=& \sigma_z \; i \sqrt{4\pi}\partial \varphi_L .
\label{z4currbos1}
\end{eqnarray}

\medskip

On the other hand, the cocycles can be directly expressed
in terms of the zero mode of the compactified Bose field $\varphi$. 
To this end,
we turn to the mode expansion of its 
chiral components $\varphi_{L,R}$:
\begin{eqnarray}
\varphi_{L} \left(z\right) &=& q_L - i \frac{p_L}{4\pi}\ln z
+ i \sum_{k \ne 0} \frac{\alpha_{L k}}{k\sqrt{4\pi}} \; z^{-k} \nonumber \\
\varphi_{R} \left(\bar z\right) &=& q_R - i \frac{p_R}{4\pi} \ln \bar z
+ i \sum_{k \ne 0} \frac{\alpha_{R k}}{k\sqrt{4\pi}} \; {\bar z}^{-k},
\label{modexp}
\end{eqnarray}
where $\alpha_{L(R) k}$ are the oscillator operators in the
quantization of the free boson and
$q_{L,R}$ and $p_{L,R}$ are
canonically conjugate operators: $[q_L,p_L] = [q_R,p_R] = i$.
Since the Bose field $\varphi$ is compactified
with radius $R = \sqrt{3/2\pi}$, the zero mode momentum
$p_{L,R}$ has the following discrete spectrum \cite{dms}
\begin{eqnarray}
p_L &=& \sqrt{\frac{2\pi}{3}} n  + \sqrt{6\pi} \; m \nonumber \\
p_R &=& \sqrt{\frac{2\pi}{3}} n  - \sqrt{6\pi} \; m ,
\label{speczeromode}
\end{eqnarray}
$n$ and $m$ being integers which correspond, respectively,
to the electric and magnetic charges associated with the
Bose field.
The cocycles can then be expressed
in terms of the zero mode $p_{L,R}$, and
the Z$_4$ parafermionic algebra is realized by
the following representation:
\begin{eqnarray}
\psi_{1L} &=& \frac{1}{\sqrt{2}}
\left(
:\exp\left(i \sqrt{6\pi} \; \varphi_L \right):
+ e^{i \sqrt{3\pi/2} \; p_{L}}
:\exp\left(-i \sqrt{6\pi} \; \varphi_L \right): \right)
\nonumber \\
\psi_{1L}^{\dagger} &=& \frac{1}{\sqrt{2}}
\left( :\exp\left(-i \sqrt{6\pi} \; \varphi_L \right):
+
:\exp\left(i \sqrt{6\pi} \; \varphi_L \right):
e^{-i \sqrt{3\pi/2} \; p_{L}} \right)
\nonumber \\
\psi_{2L} &=& -
e^{i \sqrt{3\pi/2} \;p_{L}} \sqrt{4\pi}
\; i \partial \varphi_L .
\label{z4currbos3}
\end{eqnarray}
From the spectrum of the left zero-mode momentum (\ref{speczeromode}),
one observes that the cocycle term
appearing in the above identification
takes two values: $e^{i \sqrt{3\pi/2} \;p_{L}} = \pm 1$.
It is worth noting that
the parafermionic representation of the su(2)$_4$ currents
(\ref{bosu2Nleft}), together with the formula (\ref{z4currbos3}),
provide a faithful explicit representation of the su(2)$_4$
current operators in terms of two free Bose fields.
The fact that such construction should, in principle, be possible was already
anticipated
in Ref. \cite{ZF}. An explicit representation was worked out in Ref. 
\cite{fg}. The latter construction, however, suffers from
the neglect of the cocycles, thus resulting in an incorrect
OPE for $J^{\pm}_L(z)J^{\pm}_L(w)$ (this circumstance, however, did not affect
final conclusions for the
impurity problem studied in Ref. \cite{fg} for which only
the sub-set $J^{\pm}_L(z)J^{\mp}_L(w)$ and $J^{\pm}_L(z)J^z_L(w)$
of the current OPE was actually needed). Thus 
Eqs. (\ref{bosu2Nleft}, \ref{z4currbos3})
complement the representation found in Ref. \cite{fg}.

%%%%%%%%%%%%%%%%%%%%%%%%%%%%%%%%%%%%%%%%%%%%%%%%%%%%%%%%%%%%%%%%%%%%%%%%%%%%%

\end{document}